\documentclass[
reprint,
superscriptaddress,
preprintnumbers,
 amsmath,
 amssymb,
 prc,
]{revtex4-1}

\usepackage[     bookmarks,
                 bookmarksopen = true,
                 bookmarksnumbered = true,
                 linktocpage,
                 colorlinks = true,
                 linkcolor = blue,
                 urlcolor  = blue,
                 citecolor = blue,
                 anchorcolor = green,
                 hyperindex = true,
                 hyperfigures]
        {hyperref}

\usepackage{graphicx}
\usepackage{dcolumn}
\usepackage{bm}
\usepackage{float}
\usepackage{lipsum}
\usepackage{amsmath}
\usepackage{soul}
\usepackage{color}
\usepackage{hyperref}
\usepackage{cleveref}
\usepackage{graphicx}
\usepackage{lineno}
\definecolor{dgreen}{cmyk}{1.,0.,1.,0.2}        
\definecolor{orange}{cmyk}{0.,0.353,1.,0.}    

\begin{document}

\title{Bayesian inference of event-by-event collision geometry from charged-particle multiplicity in heavy-ion collisions}

\author{Yige Huang}
\email[]{huangyige@impcas.ac.cn}
\affiliation{State Key Laboratory of Heavy Ion Science and Technology, Institute of Modern Physics, Chinese Academy of Sciences, Lanzhou 730000, China}

\author{Fu-Peng Li}
\email[]{fpli@fudan.edu.cn}
\affiliation{Key Laboratory of Nuclear Physics and Ion-beam Application (MOE) \& Institute of Modern Physics, Fudan University, Shanghai 200433, China}
\affiliation{Shanghai Research Center for Theoretical Nuclear Physics,
NSFC and Fudan University, Shanghai 200438, China}

\author{Hanwen Feng}
\email[]{hwfeng@mails.ccnu.edu.cn}
\affiliation{Key Laboratory of Quark and Lepton Physics (MOE) \& Institute of Particle Physics, Central China Normal University, Wuhan 430079, China}
\affiliation{University of Heidelberg, Heidelberg 69120, Germany}

\author{Nu Xu}
\email[]{nuxu02@gmail.com}
\affiliation{Key Laboratory of Quark and Lepton Physics (MOE) \& Institute of Particle Physics, Central China Normal University, Wuhan 430079, China}
\affiliation{State Key Laboratory of Heavy Ion Science and Technology, Institute of Modern Physics, Chinese Academy of Sciences, Lanzhou 730000, China}

\date{\today}%

\begin{abstract}
We propose the Inference-driven Participant Determination (IPD) method, a Bayesian framework for inferring event-by-event posterior distributions of the number of participants ($N_{\text{part}}$) and binary collisions ($N_{\text{coll}}$) from final-state charged-particle multiplicities in relativistic heavy-ion collisions. The joint distribution of $(N_{\text{part}}, N_{\text{coll}})$ obtained from the Monte-Carlo Glauber model is used as the prior, while negative binomial distributions calibrated to charged-particle multiplicity fluctuations define the likelihood. 
This approach replaces conventional hard-cut centrality classification with a probabilistic assignment based on $N_{\text{part}}$, making the multiplicity--geometry smearing explicit and reducing the impact of volume fluctuations on downstream observables. 
A closure test using an UrQMD-MCG hybrid model at $\sqrt{s_{NN}} = 19.6$~GeV shows that the method yields well-calibrated posterior distributions with negligible bias and improves the reconstruction of net-proton cumulants relative to conventional multiplicity-based centrality selection.

\end{abstract}

\maketitle

\section{Introduction}

In relativistic heavy-ion collisions (HICs), the initial overlap geometry, which can be quantified by the number of participant nucleons $N_\text{part}$ and binary nucleon-nucleon collisions $N_\text{coll}$, is the principal driver of bulk observables such as particle spectra, flow anisotropies, and event-by-event fluctuations\cite{STAR:2008med,Voloshin:2007af,STAR:2014egu}. 
These geometric quantities are not directly measurable; they are conventionally estimated within the framework of the Glauber model, which relates the impact parameter to the density profile of the colliding nuclei through either an optical or a Monte-Carlo formulation\cite{Miller:2007ri}. 
Accurate knowledge of this geometry on an event-by-event basis is therefore essential for connecting measured final-state observables to the properties of the quark-gluon plasma (QGP) and the nuclear equation of state (EoS).

In practice, experiments determine centrality by comparing the measured charged-particle multiplicity $N_\text{ch}$ in a reference pseudorapidity window with distributions obtained from Monte-Carlo Glauber (MCG) simulations coupled to a simple particle-production model\cite{STAR:2005gfr,Miller:2007ri,Abelev:2013qoq,Loizides:2017ack}.
Events are sorted into percentile bins (e.g., 0--5\%, 5--10\%, etc.) and each bin is mapped onto a mean $N_\text{part}$ obtained from the Glauber fit. 
This procedure, while robust for establishing global trends, is necessarily event-averaged.
More importantly, the mapping is broadened by centrality fluctuations: particle production is a stochastic process, so the final-state multiplicity fluctuates even for events with identical initial geometry. 
As a result, two events with the same $N_\text{ch}$ can originate from different $N_\text{part}$ values. 
Such smearing of the $N_\text{ch}$--$N_\text{part}$ relation makes centrality binning less precise and introduces additional fluctuations into the measurement.
These volume fluctuations have long been a major obstacle in precision experimental measurements, particularly in fluctuation analyses~\cite{Jeon:2003gk,Skokov:2012ds}. 
Suppressing such fluctuations is therefore essential, and various methods have been proposed\cite{Luo:2013bmi,Braun-Munzinger:2016yjz,Holzmann:2024wyd,Friman:2025ulh,Wang:2025fve}.

Machine-learning methods have been applied to centrality classification and impact-parameter estimation, typically by training classifiers or regressors on transport-model outputs such as UrQMD or hydrodynamic simulations~\cite{David:1994qc,OmanaKuttan:2020btb,Li:2020qqn,Li:2021plq,Basak:2023wzq,Wang:2023cac,Pang:2016vdc,Zhou:2023pti,OmanaKuttan:2022aml,OmanaKuttan:2021axp,OmanaKuttan:2020brq,OmanaKuttan:2020btb}. 
These supervised approaches inevitably inherit model-dependent assumptions from the labeled training data, which may limit their reliability when applied to experimental data where the true $N_\text{part}$ is unavailable. 
Here we address this limitation by developing the Inference-driven Participant Determination (IPD) method, a Bayesian inference framework that yields the event-by-event $(N_\text{part}, N_\text{coll})$ directly from the final-state charged-particle multiplicity $\mathbf{N}_\text{ch}$. The inference procedure relies only on Monte-Carlo Glauber geometry and fitted multiplicity likelihoods, without requiring transport- or hydrodynamic-model labels. By providing the posterior estimate of $N_\text{part}$ event by event, the method replaces the hard truncation in $N_\text{ch}$ space with a probabilistic assignment in $N_\text{part}$ space, making the smearing in the multiplicity–geometry correlation explicit rather than hidden in the hard-cut binning. We perform the validation with a UrQMD-MCG hybrid model at $\sqrt{s_{NN}} = 19.6$~GeV, an energy of particular interest in the Beam Energy Scan program of RHIC-STAR~\cite{STAR:2022hbp,STAR:2025zdq}, and demonstrate that the resulting posterior distribution quantitatively recovers the event-by-event $N_\text{part}$ distribution.

\section{Method} \label{Method}

\begin{figure*}
    \centering
    \includegraphics[width=0.65\linewidth]{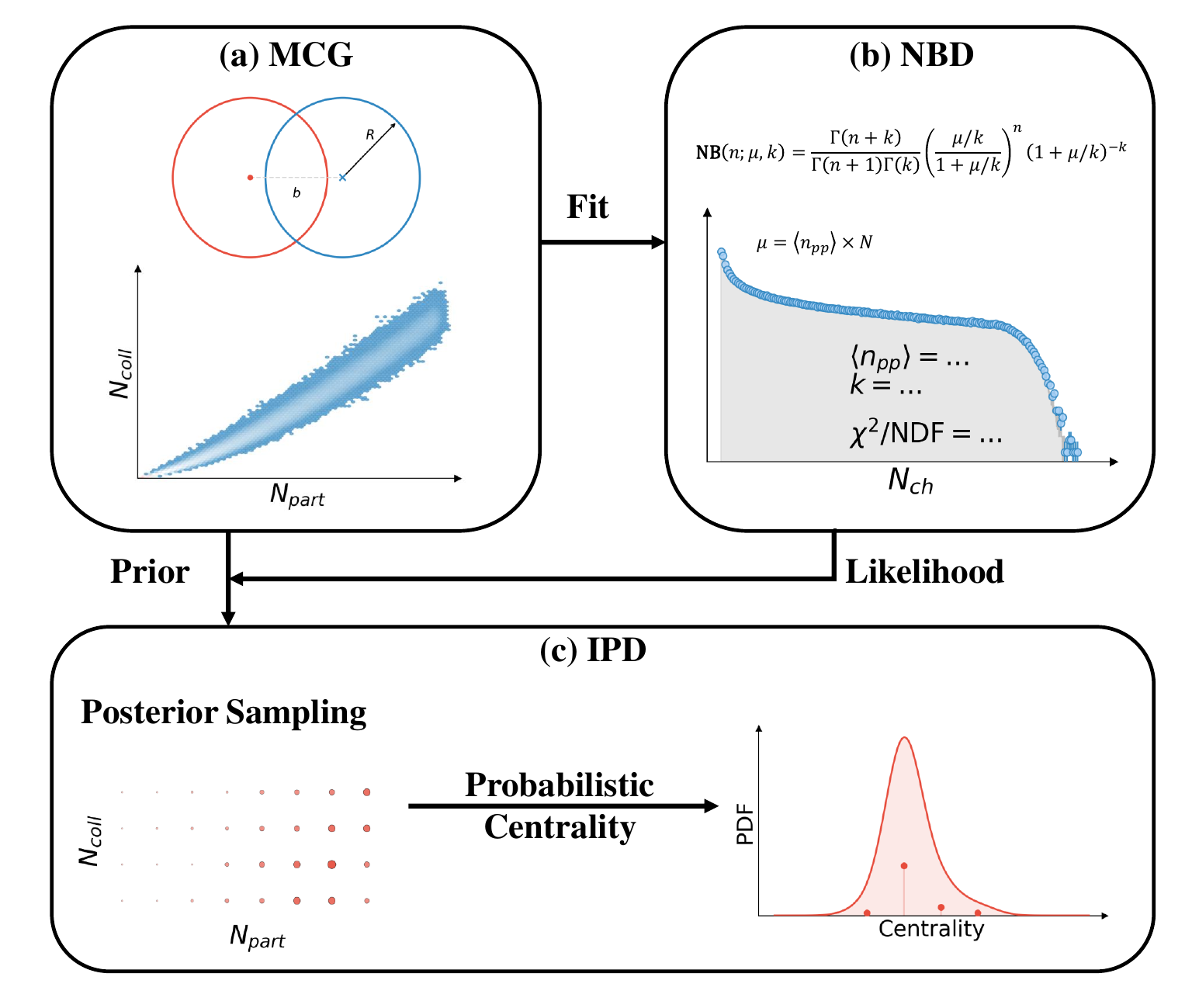}
    \caption{Schematic overview of the IPD method.
\textbf{Panel (a):} The Monte-Carlo Glauber simulation generates the joint distribution of $(N_\text{part}, N_\text{coll})$, which serves as the prior.
\textbf{Panel (b):} The negative binomial distribution describes the mapping from $(N_\text{part}, N_\text{coll})$ to $N_\text{ch}$.
The NBD parameters are determined by fitting the measured $N_\text{ch}$ distributions.
\textbf{Panel (c):} For each event, the posterior distribution of $(N_\text{part}, N_\text{coll})$ is obtained via posterior sampling from the prior weighted by the NBD likelihood.
The posterior $N_\text{part}$ samples are then projected onto predefined centrality bins, yielding a probabilistic centrality assignment for the event.}
    \label{fig:fig1}
\end{figure*}

We present a Bayesian inference framework (Inference-driven Participant Determination, IPD) that infers the event-by-event posterior distribution of $(N_\text{part}, N_\text{coll})$ from the observed charged-particle multiplicities $\mathbf{N}_\text{ch}$.
\Cref{fig:fig1} illustrates the overall workflow of the IPD method, which consists of three main components.

First, a Monte-Carlo Glauber simulation provides the joint distribution of $(N_\text{part}, N_\text{coll})$ as the prior~\cite{Miller:2007ri}.

Second, the relationship between the geometry and the final-state multiplicity is established using the conventional MCG+NBD approach~\cite{STAR:2005gfr}.
For a given $(N_\text{part}, N_\text{coll})$, an effective source number $N_\text{source}$ is obtained using a hardness parameter $x$, and the simulated charged-particle multiplicity $\hat{N}_\text{ch}$ is modeled by a negative binomial distribution with parameters $\langle n_{pp}\rangle$ and $k$ to fit the observed $N_\text{ch}$ distribution.

Third, for each event with measured $\mathbf{N}_\text{ch}$, candidate $(N_\text{part}, N_\text{coll})$ pairs are drawn from the MCG prior.
Each candidate is weighted by the NBD likelihoods, yielding a posterior distribution of $(N_\text{part}, N_\text{coll})$ for that event.
The centrality of an event is assigned probabilistically from its posterior $(N_\text{part}, N_\text{coll})$ samples: 
since the posterior can span multiple $N_\text{part}$-based centrality intervals, the probability for each class is the fraction of posterior samples whose $N_\text{part}$ falls within that interval.

\subsection{UrQMD model}

The Ultra-relativistic Quantum Molecular Dynamics (UrQMD) model~\cite{Bass:1998ca,Bleicher:1999xi} is a microscopic transport model that describes the dynamical evolution of heavy-ion collisions in terms of hadronic and string degrees of freedom,
including resonance excitation and decay, string formation and fragmentation, and hadronic rescattering.
It provides a continuous description from the initial non-equilibrium stage to the kinetic freeze-out,
and has been extensively used as a baseline for experimental measurements and methodological studies.

In this work, UrQMD simulated events are used solely for method validation, since the ground-truth $N_\text{part}$ is available for closure tests.
The IPD method itself is independent of any specific transport model and is, in principle, directly applicable to experimental data.
Details of the closure test are presented in Secs.~\ref{subsec:hybrid} and ~\ref{subsec:cumulants}.
The dataset consists of Au+Au collision events generated with UrQMD 4.0 in cascade mode, with a minimum-bias impact parameter selection of $b \sim 0$--$16$~fm, at $\sqrt{s_{NN}} = 19.6$~GeV, corresponding to representative collision energies in the BES-II program of RHIC-STAR.
Events with no interaction ($N_\text{part}<2$) are removed from the study.

\subsection{Monte-Carlo Glauber + Negative Binomial Distribution model}\label{subsec:NBD}

The MCG model is initialized with Woods–Saxon density profiles for nucleons and the inelastic nucleon–nucleon cross section $\sigma_\text{NN}$.
Through Monte-Carlo sampling of nucleon positions and impact parameters, it determines binary collisions when the transverse distance satisfies $\sqrt{b_{NN}^2} \le \sqrt{\sigma_{\text{NN}}/{\pi}}$, yielding $N_\text{part}$ and $N_\text{coll}$.
The resulting $(N_\text{part}, N_\text{coll})$ pair of an event is converted to an effective number of particle sources,
$N_\text{source} = (1-x)\cdot N_\text{part}/2 + x\cdot N_\text{coll}$,
and then combined with negative binomial distribution sampling to generate a distribution of modeled multiplicity $\hat{N}_\text{ch}$.
The negative binomial distribution is given by~\Cref{eq:nbd}.
The mean is $\mu=N_\text{source}\times \left<n_{pp}\right>$, where $\left<n_{pp}\right>$ refers to the charged-particle yields in $p+p$ collisions within specific acceptance;
the parameter $k$ controls the width of the distribution, particularly the tail, which is essential for interpreting the measured $N_\text{ch}$ distribution.
Experimentally, the detection efficiency is incorporated by binomially sampling from $n$ with detection efficiency $\epsilon$:
$\hat{N'}_\text{ch} \sim \mathrm{B}(\hat{N}_\text{ch};n,\epsilon)$.
However, as there is no detection inefficiency in the UrQMD model, $\epsilon$ is fixed to $1.0$ in this work.

\begin{equation}
\mathrm{NB}(n;\mu,k)=\frac{\Gamma(n+k)}{\Gamma(n+1)\Gamma(k)}\left(\frac{\mu/k}{1+\mu/k}\right)^{n}(1+\mu/k)^{-k}
\label{eq:nbd}
\end{equation}

The parameters in this model are determined by scanning over the parameter space and minimizing the $\chi^2/\text{ndf}$ between the modeled $\hat{N}_\text{ch}$ and the measured $N_\text{ch}$ distribution.
The centrality percentile for each event class is then defined by integrating the fitted multiplicity distribution from high to low multiplicity until a desired fraction of the total number of events is reached.
This MCG+NBD model has become a widely used framework for determining collision centrality in heavy-ion experiments~\cite{STAR:2005gfr,Abelev:2013qoq}.

For the same colliding system, different definitions of $N_\text{ch}$ correspond to distinct NBD parameter sets, each reflecting a specific category of particle production. 
In this work, four multiplicity definitions are employed. 
These are sensitive to different hadronic species, providing complementary constraints that help break the degeneracy in mapping $(N_\text{part}, N_\text{coll})$ to any single $N_\text{ch}$ observable.
These four $N_\text{ch}$ definitions are also commonly used in experimental analyses
\footnote{See the implementation in STAR's data production framework: 
\url{https://github.com/star-bnl/star-sw/blob/main/StRoot/StPicoDstMaker/StPicoUtilities.h}.}.
$N_\text{ch,1}$ is the standard charged-particle multiplicity, while $N_\text{ch,2}$, $N_\text{ch,3}$, and $N_\text{ch,4}$ count charged particles excluding pions, protons, and kaons, respectively. 
All four components of the input vector $\mathbf{N}_\text{ch} = (N_\text{ch,1}, N_\text{ch,2}, N_\text{ch,3}, N_\text{ch,4})$ are restricted to particles within the pseudorapidity range $|\eta| < 1.6$ to mimic the STAR BES-II acceptance~\cite{Yang:2017llt}.



\subsection{Inference-driven Participant Determination}

The IPD method casts the problem of extracting $(N_\text{part}, N_\text{coll})$ from $\mathbf{N}_\text{ch}$ as a Bayesian inference~\cite{DAgostini:2003bpu}.
The output for each event is a posterior estimate instead of a single point.
For an event with measured $\mathbf{N}_\text{ch}$, the posterior distribution of $(N_\text{part}, N_\text{coll})$ is given by

\begin{equation}
\begin{split}
& P(N_\text{part}, N_\text{coll} \mid \mathbf{N}_\text{ch}) \propto \\
&\quad P(\mathbf{N}_\text{ch} \mid N_\text{part}, N_\text{coll})
 \times
P(N_\text{part}, N_\text{coll}),
\end{split}
\label{eq:posterior}
\end{equation}
where the prior $P(N_\text{part}, N_\text{coll})$ is taken directly from the MCG joint distribution, and the likelihood $P(\mathbf{N}_\text{ch} \mid N_\text{part}, N_\text{coll})$ quantifies how likely the observed $\mathbf{N}_\text{ch}$ is under a given $(N_\text{part}, N_\text{coll})$ pair.
The likelihood is constructed as the product of four NBDs, each fitted independently to its corresponding $N_\text{ch}$ component:

\begin{equation}
\begin{split}
& P(\mathbf{N}_\text{ch} \mid N_\text{part}, N_\text{coll}) = \\
& \quad \prod_{i=1}^{4} \mathrm{NB}\bigl(N_{\text{ch},i}; \; \mu_i(N_\text{part}, N_\text{coll}), \; k_i(N_\text{part}, N_\text{coll})\bigr),
\end{split}
\label{eq:likelihood}
\end{equation}

with the $\mu_i\propto\langle n_{pp}\rangle_i$ and $k_i$ parameters obtained from the MCG+NBD fits.
For a single $N_\text{ch}$ component, the log-likelihood is given by

\begin{equation}
\begin{split}
& \ln \mathrm{NB}(n \mid \mu, k) =  \\
&\quad \ln \Gamma(n + k) - \ln \Gamma(n + 1) - \ln \Gamma(k) + \\
&\quad k \ln\left(\frac{k}{k + \mu}\right) + n \ln\left(\frac{\mu}{k + \mu}\right),
\end{split}
\label{eq:nbd_log}
\end{equation}
and the total log-likelihood is the sum over the four $\mathbf{N}_\text{ch}$ components.
When handling experimental data with finite detection efficiency $\epsilon$, a constant average efficiency can be absorbed into 
the NBD by rescaling $\mu \to \epsilon\mu$, eliminating the need for an explicit binomial sampling step in the likelihood calculation.

Evaluating the posterior~\eqref{eq:posterior} analytically is impractical, as the prior is a 2-dimensional discrete histogram with no closed form.
Instead, we obtain the posterior via importance sampling.
A total of $N_\text{sample}=5000$ candidate $(N_\text{part}, N_\text{coll})$ pairs are drawn from the prior distribution.
Each candidate is assigned a weight proportional to its likelihood~\eqref{eq:likelihood}.
The weights are then normalized to sum to unity, yielding a weighted sample that represents the posterior distribution of $(N_\text{part}, N_\text{coll})$ for that event.

\begin{figure}
    \centering
    \includegraphics[width=\linewidth]{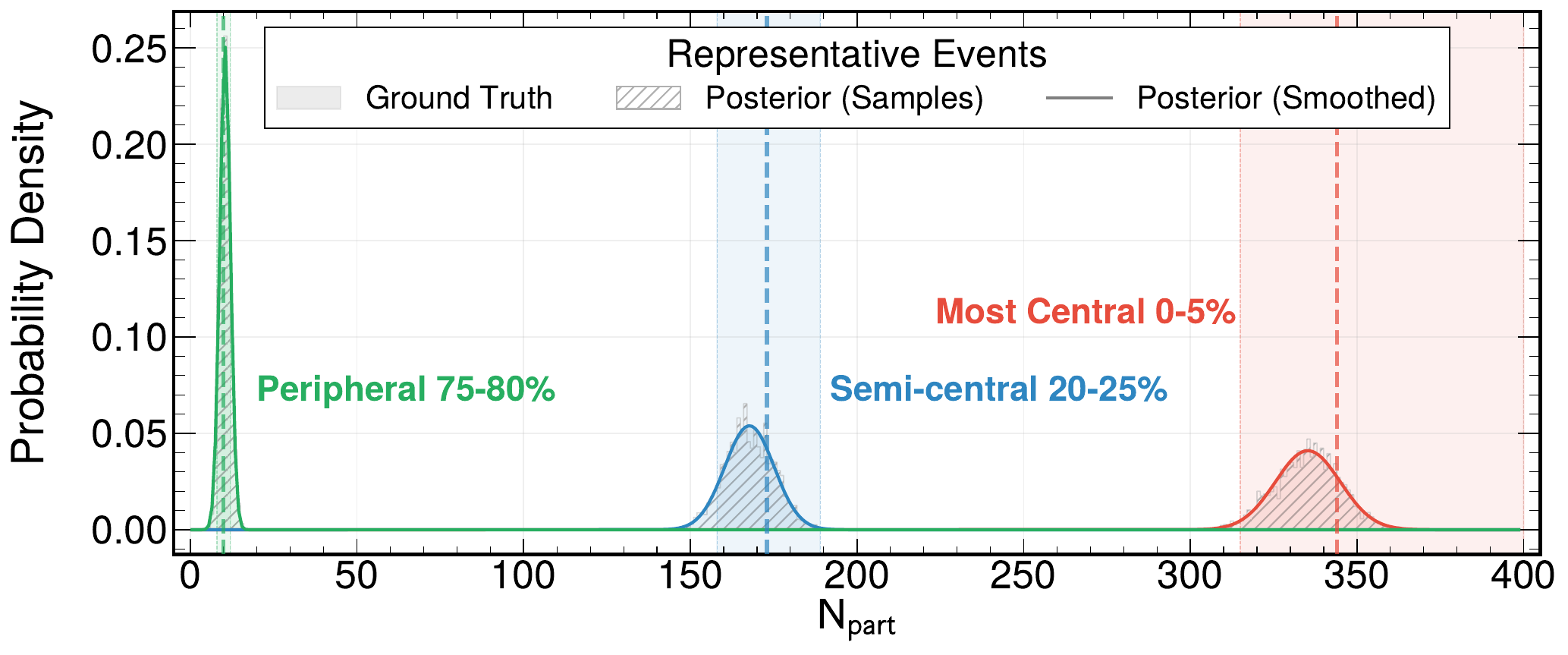}
    \caption{$N_\text{part}$ probability distributions of three representative events from the IPD method at $\sqrt{s_{NN}} = 19.6$~GeV.
    For each event, the centrality is determined by posterior $N_\text{part}$ samples scaled to a discrete PMF (hatched histogram), which is approximated by a Gaussian fit (solid curve) for visual clarity.
    The ground truth $N_\text{part}$ is shown as a vertical dashed line, and the corresponding centrality is indicated by transparent bands.}
    \label{fig:fig2}
\end{figure}

\Cref{fig:fig2} shows the posterior $N_\text{part}$ distributions for three representative events at $\sqrt{s_{NN}} = 19.6$~GeV, obtained from the IPD method.
For each event, the weighted posterior samples of $N_\text{part}$ are plotted as hatched histograms, and each distribution is smoothed by a Gaussian fit for visual clarity.
The three events are chosen to illustrate a central collision (red), a semi-central collision (blue), and a peripheral collision (green).
The true $N_\text{part}$ is shown as a dashed vertical line, with a semi-transparent band indicating the corresponding ground truth centrality bin.

In this work, the centrality bin boundaries are determined from the quantiles of the prior $N_\text{part}$ distribution.
The posterior $N_\text{part}$ samples of each event are projected onto a set of predefined $N_\text{part}$-based centrality bins. 
The probability assigned to each centrality class is simply the fraction of the total posterior weight falling into that bin, yielding a 17-dimensional (0--5\%, 5--10\%, \dots, 75--80\%, 80--100\%) probability vector per event.
More generally, the IPD method provides the full event-by-event posterior $(N_\text{part}, N_\text{coll})$ distribution, which can be used flexibly depending on the downstream analysis.

\subsection{UrQMD-MCG hybrid model}\label{subsec:hybrid}

A direct validation of the IPD method on UrQMD events is complicated by the fact that the UrQMD $N_\text{part}$ distribution is not guaranteed to match the MCG prior assumed by the IPD method.
If they differ, one cannot tell whether the bias comes from IPD method itself or from the wrong prior.

To isolate the performance of the inference framework itself, we construct a hybrid model in which the ground-truth $N_\text{part}$ is guaranteed to follow the same distribution as the MCG prior.
The model combines two independent components:
(i) the MCG joint distribution of $(N_\text{part}, N_\text{coll})$, which serves as both the prior for IPD and the ground-truth geometry;
(ii) the joint conditional distributions of proton and antiproton multiplicities given $N_\text{part}$, extracted from UrQMD data.

The generation procedure proceeds as follows.
First, a pair of $(N_\text{part}, N_\text{coll})$ is sampled from the MCG distribution.
The charged-particle multiplicities $\mathbf{N}_\text{ch}$ are then generated from the NBD likelihood with the parameters determined in ~\Cref{subsec:NBD}, ensuring that the input to IPD is consistent with its assumed likelihood.
Finally, the proton and antiproton numbers for the event are jointly drawn from the two-dimensional conditional distribution $P(p, \bar{p} \mid N_\text{part})$ constructed from UrQMD.

By construction, this hybrid model provides a controlled environment where the prior matches the underlying $N_\text{part}$ distribution, allowing the closure test to isolate the uncertainties inherent to the inference framework itself.
For each event, the true $N_\text{part}$ and the final-state particle multiplicities are recorded, enabling a direct quantitative comparison between the inferred and ground-truth results.

\subsection{Validation with net-proton cumulants}\label{subsec:cumulants}

Imperfect centrality estimation from final-state multiplicity introduces a bias in physics observables, particularly those sensitive to event classification.
Among the observables most sensitive to this bias are fluctuations of conserved-charge distributions, such as net-baryon cumulants (experimentally, net-baryon numbers are approximated by net-proton multiplicities).
These cumulants have long served as important probes in the search for the QCD critical point~\cite{Stephanov:2008qz,Asakawa:2009aj,Stephanov:2011pb}, as they are sensitive to the correlation length and are directly related to susceptibilities of conserved charges.
Volume fluctuations~\cite{Jeon:2003gk,Skokov:2012ds} are known to bias cumulant measurements, and can produce artificial signals that either wash out or mimic critical fluctuations of interest.
Given event-by-event net-proton number $N$, the cumulants up to fourth order can be expressed as $C_1 = \langle N \rangle$, $C_2 = \langle \delta N^2 \rangle$,  $C_3 = \langle \delta N^3 \rangle$, and  $C_4 = \langle \delta N^4 \rangle -3 \langle \delta N^2 \rangle^2$, where $\delta N = N - \langle N \rangle$.
Their ratios $C_2/C_1$, $C_3/C_2$, and $C_4/C_2$ are commonly used to reduce the leading volume dependence, making them sensitive to the underlying physics and less sensitive to the system size, allowing for direct comparison with model and theoretical predictions.

The IPD method infers the event-by-event $N_\text{part}$ from data, providing a more direct estimate of the initial collision geometry. 
This inference naturally reduces the volume fluctuation effects that accompany traditional centrality estimators based on multiplicity truncations.
We validate the IPD method by comparing net-proton cumulants obtained under four different centrality schemes: (i) the true $N_\text{part}$ from the hybrid model as the ground truth; (ii) IPD-based probabilistic centrality with a probability threshold applied to determine the assigned centrality bin; (iii) the conventional hard cut on $N_\text{ch,3}$; and (iv) the conventional hard cut with centrality-bin-width correction (CBWC)~\cite{Luo:2013bmi}.

\section{Results and discussion}\label{res}

Unless otherwise noted, all results presented in this section are obtained from the UrQMD-MCG hybrid model of Au+Au collisions at $\sqrt{s_{NN}} = 19.6$~GeV.
The UrQMD events are generated at this energy and serve as the source of the conditional proton and antiproton distributions, both measured within $0.4 < p_\text{T} < 2.0$~GeV/$c$ and $|y| < 0.5$.
The MCG simulation uses the standard Woods-Saxon density profile for gold nuclei ($R = 6.38$~fm, $a = 0.535$~fm) together with the NN inelastic cross section $\sigma_\text{NN} = 32.0$~mb to generate the $(N_\text{part}, N_\text{coll})$ distribution.
The hardness parameter $x=0.12$ and the detection efficiency $\epsilon=1.0$ are fixed in the NBD fitting procedure.
The NBD parameters for the four $N_\text{ch}$ definitions, obtained from the MCG+NBD fits, are listed in ~\Cref{tab:nbd_params}.
A total of $8.1\times10^7$ events are generated from the hybrid model for the closure test.

\begin{table}[htbp]
    \centering
    \caption{Parameters for $\mathbf{N}_\text{ch}$ from MCG+NBD fits.}
    \label{tab:nbd_params}
    \begin{tabular}{c|cccc}
        \hline
        ~ & $N_\text{ch,1}$ & $N_\text{ch,2}$ & $N_\text{ch,3}$ & $N_\text{ch,4}$ \\
        \hline
        $\langle n_{pp} \rangle$ & 4.48 & 0.65 & 4.18 & 4.16 \\
        k & 5.06 & 4.62 & 3.59 & 5.43 \\
        \hline
    \end{tabular}
\end{table}

\subsection{Posterior reconstruction}

The IPD method yields well-calibrated posterior uncertainties for $N_\text{part}$ inference.
The pull distribution, defined as $(\tilde{N}_\text{part} - N_\text{part}^\text{true}) / \sigma_{N_\text{part}}^\text{post}$, where $\tilde{N}_\text{part}$ is the posterior mean and $\sigma_{N_\text{part}}^\text{post}$ the posterior standard deviation of $N_\text{part}$, has a mean of $\mu_{\text{pull}} = 0.0007$ (consistent with zero) and a standard deviation of $\sigma_{\text{pull}} = 1.006$ (close to unity), indicating that the posterior uncertainty estimates are reliable and no significant systematic bias is present.

\begin{figure}
    \centering
    \includegraphics[width=\linewidth]{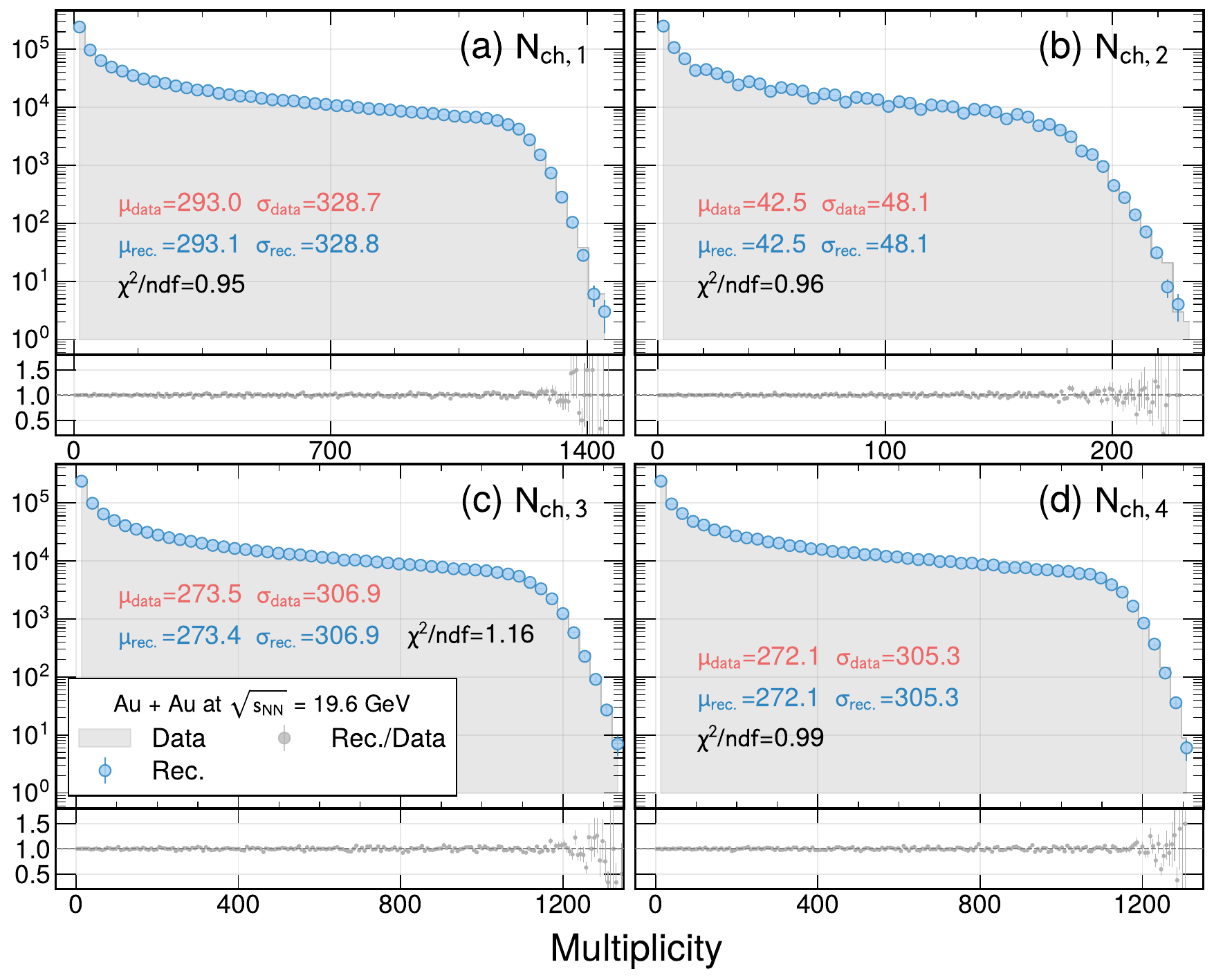}
    \caption{Comparison of $\mathbf{N}_\text{ch}$ distributions between the hybrid model (light red bands) and the IPD reconstruction (blue markers) at $\sqrt{s_\text{NN}} = 19.6$~GeV. 
    The four panels (a)–(d) correspond to $N_\text{ch,1}$ through $N_\text{ch,4}$. For each, the ratio of the reconstructed counts to the hybrid-model data is shown directly below. 
    The mean, standard deviation, and $\chi^2/\text{ndf}$ are indicated in each panel.}
    \label{fig:fig3}
\end{figure}

As a posterior predictive check, we forward-sample $\mathbf{N}_\text{ch}$ for each event by drawing from the NBD for each posterior $(N_\text{part}, N_\text{coll})$ samples. 
The resulting $\mathbf{N}_\text{ch}$ distributions are then compared with the input.
As shown in ~\Cref{fig:fig3}, the two agree well for all components of $\mathbf{N}_\text{ch}$, with all $\chi^2/\text{ndf}$ values close to unity.

\subsection{Centrality purity}

A key advantage of the IPD method is that the resulting centrality bins contain events with a purer selection in true $N_\text{part}$ compared to the conventional $N_\text{ch,3}$ hard-cut approach.
With the probabilistic centrality vector in hand, one can determine the centrality class of an event by selecting the most probable bin.
However, for events whose posterior $N_\text{part}$ distribution overlaps contiguous centrality bins, the assignment becomes ambiguous.
It is therefore appropriate to set a probability threshold $\tau$ for the classification.
By doing so, we ensure the accuracy of centrality prediction while losing some events that cannot be confidently assigned to a single centrality bin.
~\Cref{tab:frac_acc} summarizes the fraction of events retained, the exact-match accuracy, and the accuracy within one bin, as functions of $\tau$.
All events are assigned within one bin of the true centrality, with an average absolute deviation of 0.18 bins.
Balancing the retained fraction and the classification accuracy, we choose $\tau = 0.5$ as the threshold for centrality assignment in the following analysis.

\begin{table}[htbp]
    \centering
    \caption{Centrality classification performance with a probability threshold $\tau$.}
    \label{tab:frac_acc}
    \begin{tabular}{c|ccc}
        \hline
        $\tau$ & Retained fraction & Exact match & Match within 1 bin \\
        \hline
       0.4  &  100\% & 82.0\% & 100\% \\
       0.5  &  99.9\% & 82.1\% & 100\% \\
       0.6  &  88.3\% & 85.6\% & 100\% \\
       0.7  &  75.8\% & 89.0\% & 100\% \\
       0.8  &  59.9\% & 92.7\% & 100\% \\
       0.9  &  38.7\% & 96.8\% & 100\% \\
        \hline
    \end{tabular}
\end{table}

\begin{figure}[hbtp]
    \centering
    \includegraphics[width=\linewidth]{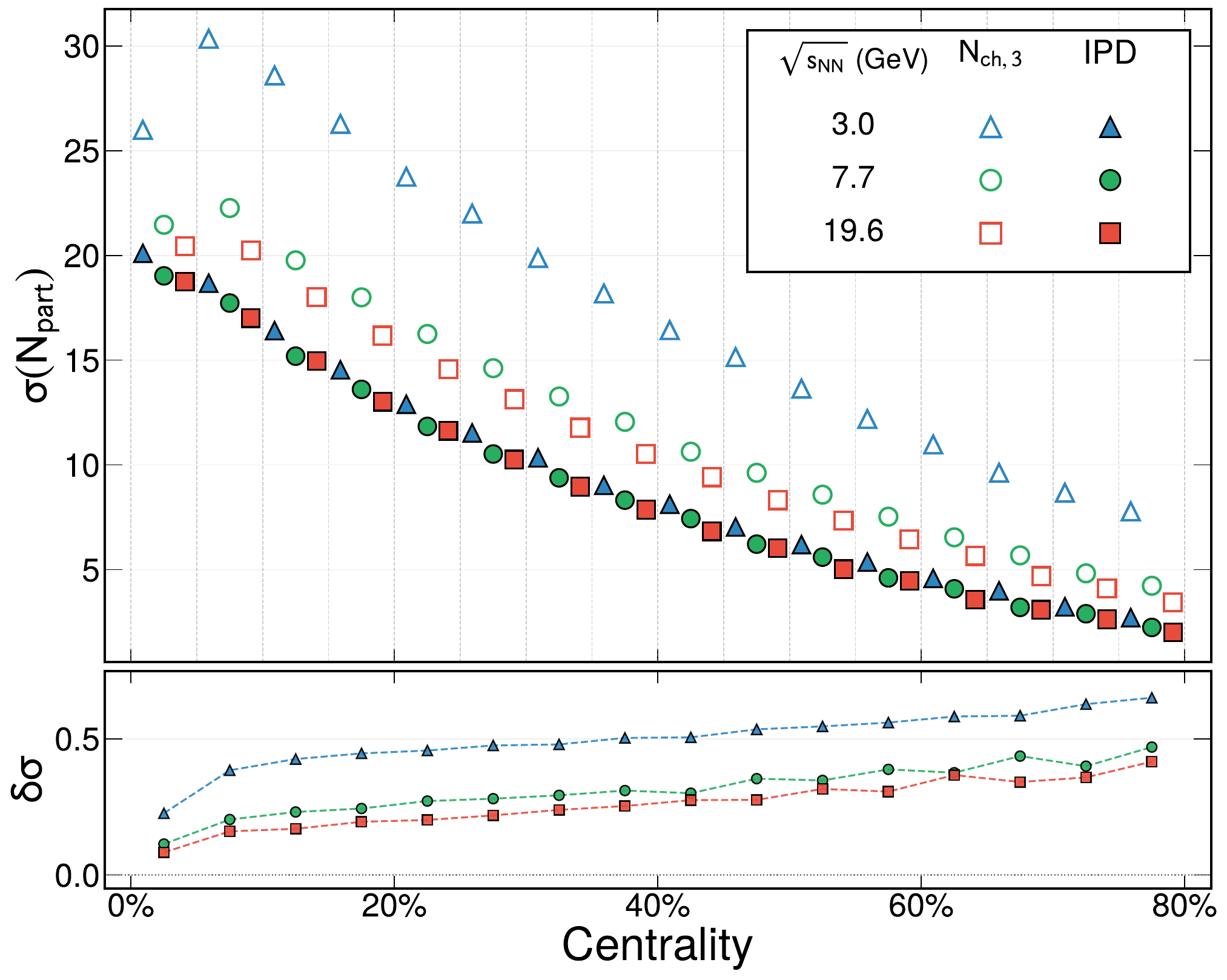}
    \caption{RMS width $\sigma$ of the $N_\text{part}$ distribution as a function of centrality, comparing $N_\text{ch,3}$-based (open markers) and IPD-based (solid markers) centrality selection. 
    Results are shown for $\sqrt{s_\text{NN}} = 3.0$, 7.7, and 19.6~GeV. \
    Across all energies, IPD yields a narrower $N_\text{part}$ distribution in every centrality bin.
    The lower panel shows $\delta\sigma$ as a function of centrality for each energy.}
    \label{fig:fig4}
\end{figure}

~\Cref{fig:fig4} shows the RMS width $\sigma$ of the $N_\text{part}$ distribution as a function of centrality for $N_\text{ch,3}$-based and IPD-based centrality selection, compared across three collision energies: $\sqrt{s_\text{NN}} = 3.0$, 7.7, and 19.6~GeV. 
For this comparison, the same hybrid model analysis framework is applied at all three energies (9M events for 3.0 and 7.7 GeV). 
At a given energy, both the $N_\text{ch,3}$-based and the IPD-based $\sigma(N_\text{part})$ decrease from central to peripheral collisions. 
Across energies, $\sigma(N_\text{part})$ decreases monotonically with increasing $\sqrt{s_\text{NN}}$, for both centrality definitions. 
The corresponding $\delta\sigma \equiv 1 - \sigma_\text{IPD}/\sigma_{N_\text{ch,3}}$ varies systematically with both collision energy and centrality, being largest at 3.0~GeV and in peripheral collisions, where the $N_\text{ch,3}$-based classification is least precise.
At all three energies, IPD yields a narrower $N_{\rm part}$ distribution across the full centrality range, confirming that the improvement in centrality purity is robust.

\begin{figure}[hbtp]
    \centering
    \includegraphics[width=\linewidth]{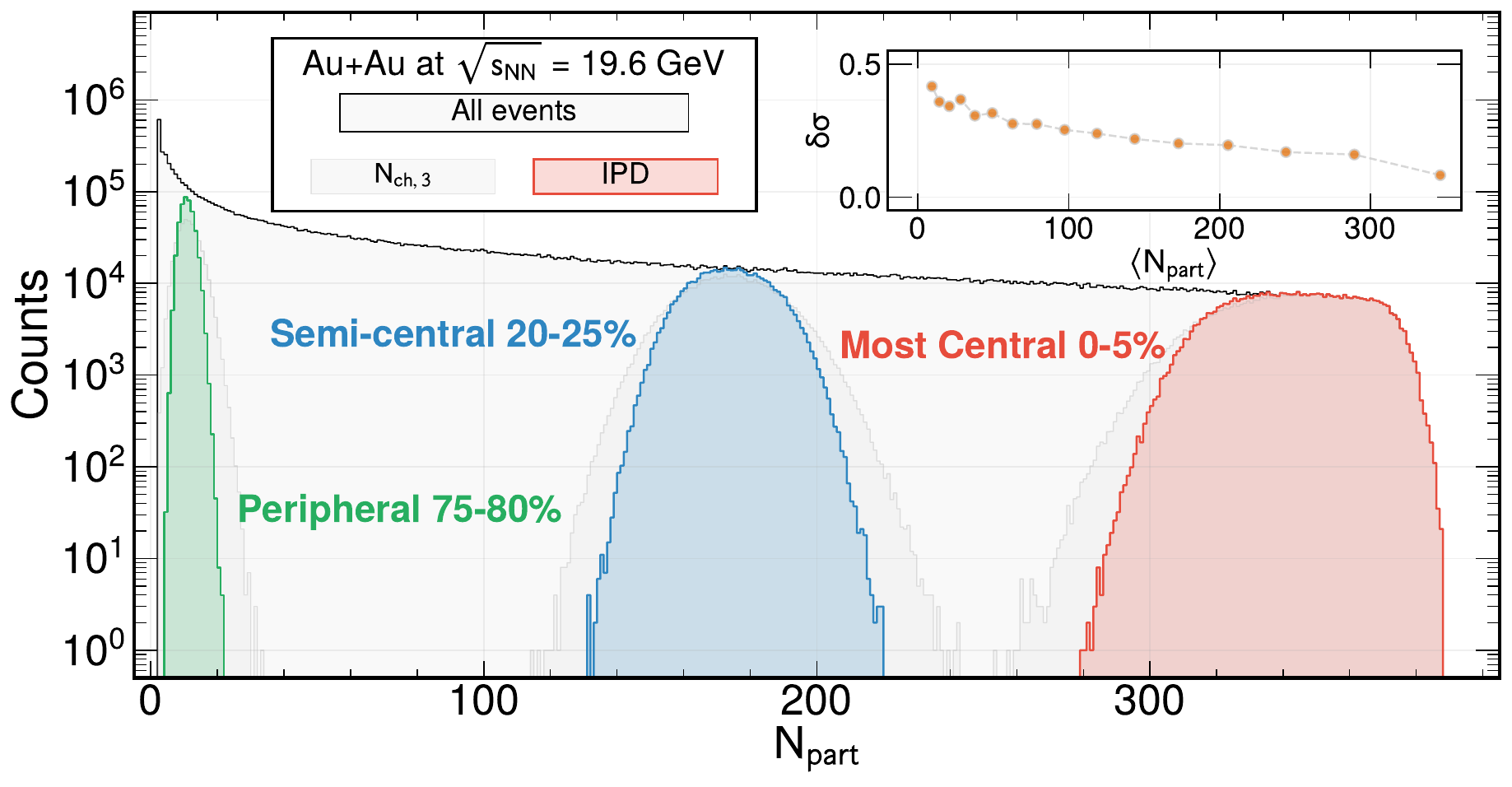}
    \caption{Comparison of $N_\text{part}$ distributions in 3 centrality classes from IPD (colored, $\tau=0.5$) and $N_\text{ch,3}$-based (gray) binning at $\sqrt{s_\text{NN}} = 19.6$~GeV.
    The black line shows the total $N_\text{part}$ distribution of all events.
    The IPD method yields narrower $N_\text{part}$ distributions, indicating improved centrality purity.
    The inset shows $\delta\sigma$ as a function of $\langle N_\text{part}\rangle$ for each centrality class.}
    \label{fig:fig5}
\end{figure}

~\Cref{fig:fig5} shows the $N_\text{part}$ distributions for IPD-based and $N_\text{ch,3}$-based centrality at $\sqrt{s_\text{NN}} = 19.6$~GeV. 
Quantitatively, the RMS width $\sigma$ is reduced by 8.3\% in the most central bin and up to 41.6\% in the peripheral bin. 
As shown in the inset, $\delta\sigma$ grows systematically toward peripheral collisions, following a nearly monotonic trend with only minor deviations in the moderately peripheral region.

\subsection{Net-proton cumulant closure test}

We now turn to the net-proton cumulants as a critical test of the IPD method.
They, in particular higher orders, are known to be sensitive to the quality of centrality determination, and serve as a benchmark for evaluating the IPD performance against the ground truth and the conventional approaches.
The cumulants are calculated for four centrality schemes — true $N_\text{part}$ (as Ground Truth), IPD with $\tau=0.5$, $N_\text{ch,3}$ hard cut, and $N_\text{ch,3}$ hard cut with centrality bin width correction (CBWC)~\cite{Luo:2013bmi}.
The statistical uncertainties are estimated by bootstrap resampling with 500 resamples.

\begin{figure}
    \centering
    \includegraphics[width=\linewidth]{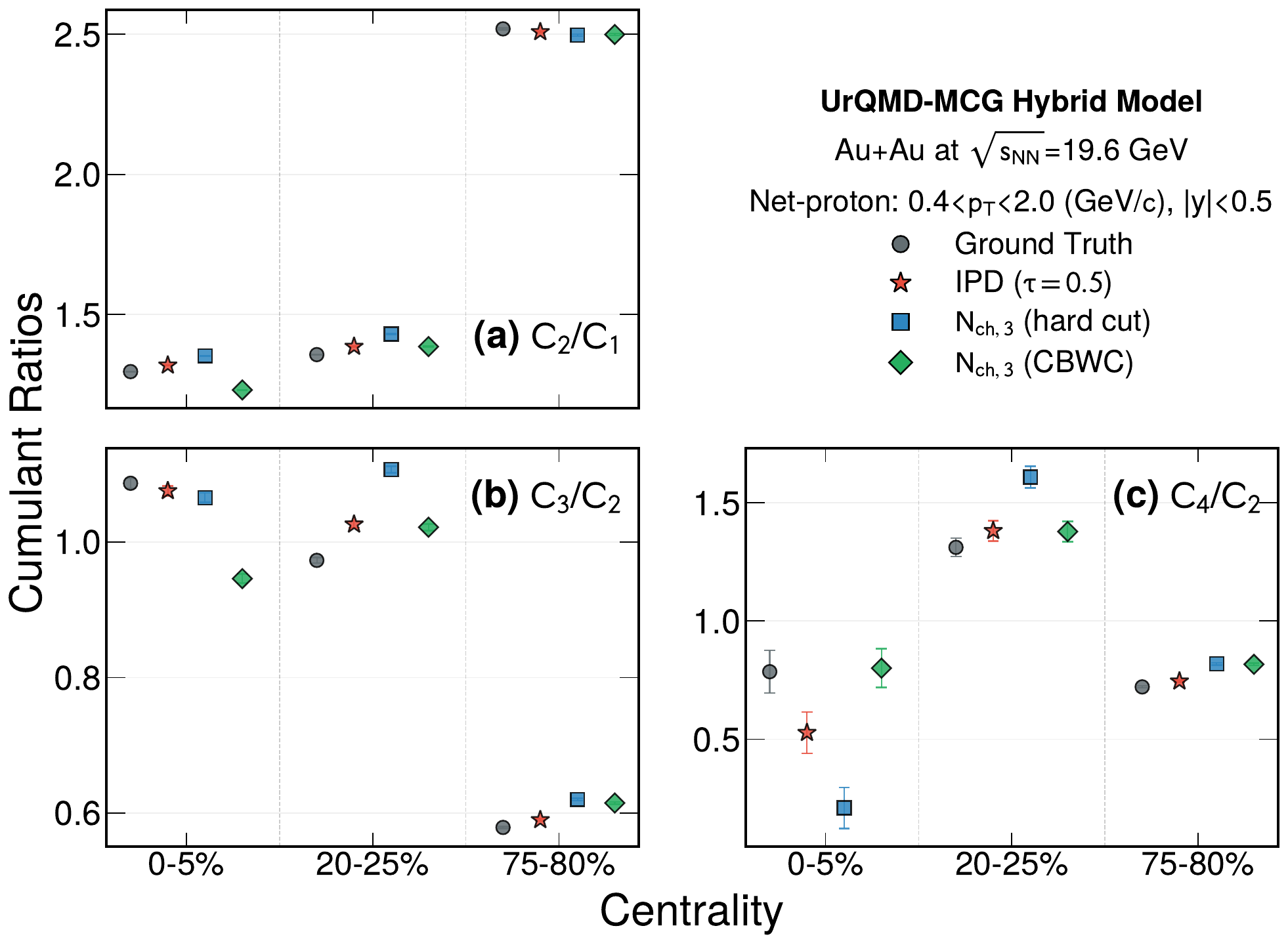}
    \caption{Net-proton cumulant ratios $C_2/C_1$, $C_3/C_2$, and $C_4/C_2$ for three centrality classes at $\sqrt{s_{NN}} = 19.6$~GeV. 
    Within each centrality class, four results are shown: true $N_\text{part}$ (Ground Truth, gray), IPD with $\tau=0.5$ (red), conventional $N_\text{ch,3}$ hard cut (blue), and $N_\text{ch,3}$ hard cut with CBWC (green). 
    The protons and antiprotons are measured within $0.4 < p_\text{T} < 2.0$~GeV/$c$ and $|y| < 0.5$. 
    Error bars represent statistical uncertainties from 500 bootstrap resamples.}
    \label{fig:fig6}
\end{figure}

\Cref{fig:fig6} shows the cumulant ratios $C_2/C_1$, $C_3/C_2$, and $C_4/C_2$ for three representative centrality classes (0--5\%,20--25\%, and 75--80\%), each compared across the centrality schemes.
The $N_{\text{ch,3}}$-based hard cut does not incorporate any treatment for volume fluctuation and is shown as a reference. 
We focus the discussion on the IPD and CBWC methods.

For $C_2/C_1$, both IPD and CBWC reproduce the ground truth to within 5\% relative deviation across all three centrality classes.
For $C_3/C_2$, IPD and CBWC perform comparably in semi-central collisions, while IPD is generally closer to the ground truth in central and peripheral collisions.
For $C_4/C_2$, the deviations are quoted in units of the combined statistical uncertainty $\sigma$. 
In the central bin CBWC agrees with the ground truth to within $0.1\sigma$, while IPD deviates by $2.0\sigma$ below.
In peripheral collisions, where volume fluctuations play a stronger role, IPD deviates from the ground truth by $4.1\sigma$, whereas CBWC shows a $15.1\sigma$ discrepancy.
In semi-central collisions the two methods yield comparable deviations, similarly to the lower-order ratios.
We have verified that the cumulant results are consistent within uncertainties for $\tau \in [0.5, 0.7]$, demonstrating the robustness of the method to this choice.
Overall, IPD provides the closest agreement with the ground truth for the majority of cumulant ratios and centrality bins, with the notable exception of the central $C_4/C_2$ where CBWC performs best. 
In peripheral collisions, where the impact of volume fluctuations is greatest, IPD shows a clear advantage over CBWC across all three cumulant ratios. 
This closure test suggests that the improved centrality purity by IPD leads to a measurable reduction of systematic bias in cumulant observables.

\section{Summary and outlook}\label{sum}

In this paper, we have developed the Inference-driven Participant Determination (IPD) method, a Bayesian framework that infers the event-by-event $(N_\text{part}, N_\text{coll})$ distribution from charged-particle multiplicities using a Monte-Carlo Glauber prior and a negative binomial likelihood. 
It provides a probabilistic centrality assignment for each event, enabling flexible downstream analyses. 

The IPD method is validated through a self-consistent closure test using an UrQMD-MCG hybrid model in Au + Au collisions at $\sqrt{s_\text{NN}}=19.6$ ~GeV, which supplies both the MCG geometry (as the prior) and the UrQMD-derived proton and antiproton distributions, ensuring that the prior matches the ground-truth $(N_\text{part}, N_\text{coll})$ distribution.
Net-proton cumulants, which are sensitive to centrality classification, serve as a benchmark for the validation.
The results show that IPD-based centrality yields narrower true $N_\text{part}$ distributions within each centrality bin, with the RMS width $\sigma$ reduced by 8\% to 42\% ($\delta\sigma \equiv 1 - \sigma_\text{IPD}/\sigma_{N_\text{ch,3}}$) from central to peripheral collisions, and the corresponding net-proton cumulants are closer to the ground truth across most centrality classes and cumulant orders, especially in peripheral collisions where volume fluctuations are strong, compared to conventional $N_{\text{ch,3}}$-based approaches.

Unlike methods that rely on transport-model training, the IPD method requires only the MCG geometry and the measured $\mathbf{N}_\text{ch}$ distributions, making it directly applicable to experimental data without transport-model-dependent assumptions.
The method has been tested at $\sqrt{s_\text{NN}} = 3.0$, 7.7, and 19.6~GeV, covering the STAR BES-II energies with upgraded detector capabilities, and yields consistent improvements across all three energies.
The IPD framework is general and can be applied to any collision system and energy where the MCG+NBD description of $\mathbf{N}_\text{ch}$ is valid, providing a unified centrality classification for a wide range of observables.
It can be readily adopted by a variety of heavy-ion experiments, from the STAR Beam Energy Scan program to future facilities such as CBM~\cite{CBM:2016kpk} and CEE~\cite{Lu:2016htm}.

\section{Acknowledgments}

Y.G. Huang acknowledges support from Gansu Provincial Natural Science Foundation Postdoctoral Special Project, and the National Key Research and Development Program of China under Grant No. 2024YFA1610700.
F.P. Li acknowledges support from the National Natural Science Foundation of China (NSFC) under Grant Nos. 12325507, 12547102, and 12147101, from the National Key Research and Development Program of China under Grant No. 2022YFA1604900, and in part from the China Postdoctoral Science Foundation under Grant No. 2025M783370.

\section{Declaration of generative AI and AI-assisted technologies in the writing process}

During the preparation of this work, the author(s) used deepseek in order to polish the language. 
The author(s) reviewed and edited the output as needed and take full responsibility for the content of the published article. 

\bibliography{ref}

\appendix

\end{document}